\documentclass[twocolumn,prl]{revtex4}

\usepackage{graphicx}
\usepackage{dcolumn}
\usepackage{bm}

\begin{document}

\title{Disproof of ``Disproof of Bell's Theorem by Clifford Algebra Valued Local Variables"}

\author{Tung Ten Yong}
\email{tytung2020@hotmail.com}

\begin{abstract}
A key mathematical error is pointed out regarding the paper "Disproof of Bell's
Theorem by Clifford Algebra Valued Local Variables" by Joy
Christian. 
\end{abstract}

\maketitle

\section{\label{sec:level1}Original Comment}

In a series of papers (\footnote{``Disproof of Bell¡¯s Theorem by
Clifford Algebra Valued Local Variables", Joy Christian,
quant-ph/0703179}, \footnote{``Disproof of Bell's Theorem: Reply to
Critics", Joy Christian, quant-ph/0703244} and \footnote{``Disproof
of Bell's Theorem: Further Consolidations", Joy Christian,
quant-ph/0707.1333}), Joy Christian challenges the mainstream view
by showing that a local hidden variable model can violate the Bell's
inequalities. The novel feature of the model is that instead of the
usual real number, the complete state of quantum systems is
described by geometric entities called trivectors, and the spin
observables are described by bivectors. Both are elements in
geometric algebra.

The model and the relevant geometric algebra notions will not be
introduced here \footnote{See, eg. C. Doran, A.Lasenby,
\emph{Geometric Algebra for Physicists}, Cambridge University Press
(2003).}, but will only point out a quite elementary mathematical
error that invalidates the whole argument: The expectation value in
the equation (19) of paper [1] is not a valid expectation value at
all. An expectation value function is a function with a real
codomain \footnote{See, eg. A. Gut, \emph{Probability: A Graduate
Course}, Springer (2007).}, but the codomain of the function in
equation (19):
\begin{eqnarray} \int(\bm{\mu}\cdot \bm{a})(\bm{\mu}\cdot
\bm{b}) d\bm{\rho}(\bm{\mu})
\end{eqnarray}
is not the space of scalar, but the larger space of multivectors.
This is seen clearly from the equation (17) of the paper:
\begin{eqnarray} (\bm{\mu}\cdot
\bm{a})(\bm{\mu}\cdot \bm{b})=
-\bm{a}\cdot\bm{b}-\bm{\mu}\cdot(\bm{a}\times\bm{b})
\end{eqnarray}
The second term on right hand side is a bivector, as seen from the
equation (15) of the paper. Therefore the expectation value (1)
above is a (directed) integration of the sum of scalar and bivector.
The result is not (just) a scalar. In fact, this can be clearly seen from equation (18) of
the paper, the correct answer of which should be a null vector (cancelling of two vectors is still a vector, not a scalar. An 
elementary fact about vector space.)\footnote{Regarding Equation (19), this can clearly be seen by letting vectors $\bm{a}$ and $\bm{b}$ to be orthogonal 
to each other. Then we will have $-\int\bm{\mu}\cdot(\bm{a}\times\bm{b})d\bm{\rho}(\bm{\mu})$, where
inside the integration is the product of a bivector and a trivector (measure), 
which is just a vector. In the case of isotropic distribution assumed in the paper,
we get a null vector, not a zero scalar.}
\footnote{It would be question begging if the distribution is
required to be isotropic in order to reproduce the quantum
correlation results.}. As a consequence, the correlations calculated
from such `expectation functions' do not have the usual statistical
meanings, and therefore their violations of Bell's inequalities is
meaningless and do not have the significance as claimed by the
author.

In order to obtain a valid correlation function from the two
bivectors in (1) above, we should have a map $f$ such that
:
\begin{eqnarray} f: (\bm{\mu}\cdot \bm{a},\bm{\mu}\cdot \bm{b}, d\bm{\rho}(\bm{\mu}))
\rightarrow \mathbf{R}
\end{eqnarray}
before summation. However, there is no unique way to do so. The simplest way is by
letting each $\bm{\mu}\cdot \bm{a}$, $\bm{\mu}\cdot \bm{b}$ mapped into scalars,
and $d\bm{\rho}(\bm{\mu})$ mapped into a scalar measure, and then take the product. 
But this will then yield correlation functions that obey Bell's inequalities. Another
possible way is to first perform some operations between the three
multivectors and then map the resulting quantity (which need not be a
scalar) into a scalar. In this case however one needs to justify
what kind of operations are good operations, i.e. such that the
scalar that is finally obtained can be taken to depict the
correlation between the two bivectors. Since Christian's paper did
not do so, his argument is based on unjustifiable
assumptions.

It should be noted that the error pointed out here is different from
those pointed out by Philippe Grangier in \footnote {`` `Disproof of
Bell's theorem': more critics'', Philippe Grangier,
arXiv:0707.2223}, where he did not question the validity of the
correlation functions in Christian's paper, but instead refute the
paper's use of bivectors to represent measurement results. While his
refutation is more on the conceptual side, the error pointed out in
this paper is much more elementary and decisive.

The first comment on Christian's paper, \footnote {``Comment on:
Disproof of Bell's Theorem by Clifford Algebra Valued Local
Variables'', Marcin Pawlowski, quant-ph/0703218} by Marcin
Pawlowski, is also different from the current paper. That comment
seems to say that in Clifford-algebra-valued hidden variable theory
it is unable to derive Bell's inequalities. This is not true since
they are indeed derivable, as is explicitly shown in [2].

\section{\label{sec:level1} Updated Comment: Reply to arXiv:quant-ph/0703244v9}

In order to avoid further misunderstandings, this new section will spell out as clearly as possible 
the relevant mathematical error made in Joy Christian's papers, before replying to his quant-ph/0703244v9.

a) The expectation value in his theory is a function with codomain the total geometric algebra of 3-dimensional
Euclidean space, $\mathcal{G}_{3}$ (notation see reference 4), not the grade-0 subspace (scalar) of it, as can be seen from the way he constructs the function. 
Therefore his `expectation value' is actually a linear combination of elements from subspaces of different grades, i.e. a linear combination of some
 scalar, vector, bivector and trivector. For example, codomain of the function in Equation (18) is grade-1 subspace (vector space) of $\mathcal{G}_{3}$, while that in Equation (19) is the space span by grade-1 and grade-3 (trivector) subspaces of $\mathcal{G}_{3}$. 

If this point is understood, then the error is obvious and no furthur arguments need be given.
But if one cannot see this, then furthur clarifications can be given as follows:

b) If the expectation value is a valid one (a function with scalar codomain), then it must do so \emph{whether the distribution is
isotropic or not}. When the distribution is not isotropic, the expectation value in his papers will in general contain a 
non-zero grade-1 component (vector).
If the distribution is isotropic as assumed in his papers, the error still persists (but is very easy to be overlooked) 
because now the grade-1 element is a zero element of that subspace (i.e. a zero vector). 
Thus the reason to point out the non-isotropic case is merely to show more clearly the codomain of the function.

None of his replies answers the objections made above.

c) But if the above is too convoluted then just focus on the term in the third line in Equation (19). 
This should be a trivector and not a scalar, because it is the consequence of directed integration (with trivector measure) of a scalar $-\bm{a}\cdot\bm{b}$. 
So even if one can ignore the vanishing terms we still don't get a scalar out of the function. In fact this already shows that
the trivector valued measure is not a valid probability measure at all, because they do not normalized to a scalar unity.

Now, regarding his reply (in quant-ph/0703244v9) on null vector or zero vector, he totally misses my point.
By a null vector of course I meant a vector with zero magnitude, which is just what he mean by zero vector. As he also said in that reply,
a zero vector is \emph{a vector} with all components zero value. But this is exactly my objection: it is not a scalar!

\section{\label{sec:level1} Updated Comment: Reply to arXiv:quant-ph/0703244v10}

The preprint quant-ph/0703244v10 avoids answering to my main objection (a) above (in fact, it avoids answering all of the objections above). 
He seems to continue to missunderstand my objections and still think that the main objection hinges on whether the distribution is isotropic or not. 
Let me emphasize again that this is not so. The main problem is that his probabilistic functions are not our usual numbered probabilistic functions. 
The fact that Clifford algebra blurrs the distinction between scalars and vectors precisely implies that they cannot be used in probabilistic theory in the way that he employs them 
(none of my sentence says that such blurring is an ``error''). The fact that in such framework it makes no sense to speak of a scalar codomain implies precisely 
that such framework cannot be employed here (at least in the way that he does it). That the algebra blurrs the distinction cannot be taken to mean that we are now allowed to 
have a theory that will give, eg. a multivector correlation function, a vectorial average value etc. We need to know what such functions means before we can understand them in the 
way we understand our usual number valued statistical functions. This is something that is not obvious and is one that \emph{needs to be justified}, which Joy Christian does not do and simply assumes 
that they already are. Whatever a local realist's goals are, he needs to use legitimate statistical functions and probabilistic measures. One cannot simply use non-statistical functions to violate 
Bell's inequalities and then claim that this disproves Bell's claims. 

If this is understood and agreed upon, then it is baffling why he still does not see the problem. In fact this is already clear from his statement: `the word ``scalar'' within this framework simply means \emph{a multivector} whose all but 
one components are strictly vanishing.'(emphasis by me). This immediately means that all his `statistical functions' (expectation, correlation) are actually multivectors and subjects to my objection above.
(His comment on my remark about zero vector and zero scalar is not directly related to my objection, so my reply is put in a footnote \footnote{Here he made a double error and also misunderstood my sentence. First, there is of course a difference between 
a zero element in grade-1 subspace and a zero element in grade-0 subspace. They are identity elements in the respective subspaces relative to additions within the spaces. Secondly, when he mentions `zero scalar' he is actually 
referring to the identity element of the total $\mathcal{G}_{3}$. There is a conceptual difference here. In this context, my remark was actually referring to the identity elements of the subspaces.}.)

Lastly, it is in fact his mistake that he takes the third line of his Equation (19), a trivector measure integration of a scalar to be a scalar. Addition is closed \emph{within} grade-3 subspace of $\mathcal{G}_{3}$ 
(an elementary fact about Geometric algebra): summation of trivectors gives a trivector. So his use of trivector measure as probabilistic measure is totally erroneous (It is urged that Joy Christian should explicitly point out the error of this instead of avoiding to give any answers, which only shows that
he does not want to admit the elementary mistakes that he made).

\section{\label{sec:level1} Updated Comment: Reply to arXiv:quant-ph/0703244v11}

This will be my final comment to Joy Christian's papers and his replies. 

It seems that from his reply he yet again failed to notice, and continues to misunderstand, the central objection to his model.
Even if the \emph{mathematical} theory of probability can be 
formulated in a more generally way such that the codomain of functions is not necessarily real number, this is really unrelated to the main objection. 
First, the relevant functions in such formulations will eventually have to be mapped into real numbers if one is to connect with experimental results. 
In fact such generalized `statistical functions' do not has the direct meaning as their original and more limited counterparts, and has to be interpreted before can be understood in the usual sense. 
Second, Bell's theorem is based on the usual probability theory, the significance and meaning that we usually attaches to the theorem relies on the use of it, 
because in experiments testing the inequalities we only encounters scalar frequencies and records numbered measurement results. 
It is within such contexts that I claimed that his `statistical functions' are not valid (i.e. they are not valid functions to be compared directly with the corresponding functions obtained from 
experiment results, which is something that von Neumann and Segal did not do).

Therefore it is of no use that one has a model that do not have a strictly scalar output if one wants to compare this to the
usual Bell's inequalities, and one cannot attach the same meaning to the violation of its version of `Bell's inequalities' as
to the violation by a scalar-valued-observable hidden variable theory. Justification is needed if one wants to claim that it does has the same meaning.

Besides the above conceptual problem, there is also mathematical error. In arXiv:quant-ph/0703244v11, he claims that his model gives functions (linear functionals) with codomain effectively the real $[-1,1]$.
But from his equations it is obvious that the functionals, curiously, can only obtain the zero multivector, and cannot obtain other scalar values within the domain. 
This is not so curious if one is reminded that the correct codomain that one gets for these functionals (when one varies the distribution) is the non-scalar linear subspace
of $\mathcal{G}_{3}$ (see (a) above). So his (36) and (37) are both wrong and very misleading.

Lastly, there is no error in my comment that his directed measure is not the usual valid probability measure, since the former normalizes to
a trivector (as in his equation (39)) but the latter normalizes to a scalar one. 

\end{document}